
\input amstex
\documentstyle{amsppt}
\def\lowbar#1{\lower 0.5em \hbox{$\bigr|_{#1}$}}
\rightheadtext{B\"ACKLUND TRANSFORMATION AND \dots}
\topmatter
\title
B\"ACKLUND TRANSFORMATION AND THE CONSTRUCTION OF THE
INTEGRABLE BOUNDARY-VALUE PROBLEM FOR THE EQUATION
$u_{xx}-u_{tt}=e^u-e^{-2u}$.
\endtitle
\author
Sharipov R.A. and Yamilov R.I.
\endauthor
\abstract
B\"acklund transformation for the Bullough-Dodd-Jiber-Shabat
equation $u_{xx}-u_{tt}=e^u-e^{-2u}$ is found. The
construction of integrable boundary condition for this equation
together with the algebro-geometric solutions satisfying it are
suggested.
\endabstract
\endtopmatter
\head
     1. Introduction.
\endhead
     Most features of integrability for nonlinear differential
equations are considered on a base of the inverse scattering
method for solving the initial-value Cauchy problem. They are
not specific for the boundary-value problems for those
equations. Exception are the special forms of boundary
conditions preserving the integrability. Nontrivial boundary
conditions of such kind was first considered in \cite{1} by
means of $r$-matrix approach. In \cite{2--5} some methods of
solving the integrable boundary-value problems for nonlinear
Schrodinger equation and for Sine-Gordon equation are developed.
In this paper we suggest the scheme for finding the proper
boundary condition of the form
$$
R(e^u,u_x,u_t,u_{xt})\lowbar{x=0}=0\tag1.1
$$
for the Bullough-Dodd-Jiber-Shabat equation
$$
u_{xx}-u_{tt}=e^u-e^{-2u}\tag1.2
$$
together with the algebro-geometric solutions of the
boundary-value problem arisen. We also suggest the B\"acklund
transformation for this equation different from that of
\cite{6}. \par
     First in section 2 we consider the mirror symmetric pairs of
the algebro-geometric solutions $v(x,t)$ and $\hat v(x,t)$ of the
equation \thetag{1.2}
$$
\hat v(x,t)=v(-x,t)\tag1.3
$$
and the way how to construct them. Then in section 3 we set an
additional restriction for such pairs taking $v$ and $\hat v$
to be bound with the B\"acklund transformations
$$
v(x,t)\longrightarrow u(x,t)\longleftarrow \hat v(x,t)
\tag1.4
$$
via the virtual one-soliton solution $u(x,t)$ of the equation
\thetag{1.2}. This gives the opportunity to get several
differential relationships between $v(x,t)$ and $u(x,t)$ on a
boundary $x=0$. The number of these relationships is large
enough to exclude $u(x,t)$ with its derivatives and to get
the boundary condition of the form \thetag{1.1} for the solution
$v(x,t)$. In section 4 we managed only to reduce the exclusion
problem to the system of four polynomials with three extra
variables to be excluded. Being algorithmically solvable this
problem though may require the use of computer methods to get
the explicit form of polynomial $R$ in \thetag{1.1}.\par
     We should note that for a long time the equation
\thetag{1.2} was thought to have no auto-B\"acklund transformation
(see \cite{9--11}). In those papers authors sought B\"acklund
transformation in the class of differential relationships
depending on the first order derivatives of $u$. Our B\"acklund
transformation contain the second order derivatives, therefore
it is out of range of theorems proved there. \par
\head
2. Mirror-symmetric finite-gap solutions.
\endhead
     Let $\Gamma$ be an even genus Riemann surface realized as a
three-fold ramified covering of the complex $\lambda$-plane with
the branching points at $\lambda=0$, at infinity and at the
points $\pm\lambda_1,\dots,\pm\lambda_g$ on a real axis. Riemann
surfaces of such kind are connected with the class of the
algebro-geometric or so called finite-gap solutions of the
equation \thetag{1.2} being the complexified spectra of Lax operators
for such solutions (see \cite{7, 8}). In order to have the same
notations with \cite{7, 8} now in sections 2 and 3 we consider
the equation \thetag{1.2} in "light cone" variables where it has the
form
$$
u_{xt}=e^u-e^{-2u}\tag2.1
$$
Let us consider two finite gap solutions $v(x,t)$ and $\hat
v(x,t)$ of the equation \thetag{2.1} constructed  with the help
of two vectorial Baker-Achiezer functions $e(x,t,P)$ and $\hat
e(x,t,P)$ depending on $P\in\Gamma$. Let $D$ and $\hat D$ be the
divisors of poles of $e(x,t,P)$ and $\hat e(x,t,P)$
respectively. For $v(x,t)$ and $\hat v(x,t)$ to obey the
relationship
$$
\hat v(x,t)=v(-t/\varkappa^2,-\varkappa^2 x)\tag2.2
$$
(compare with \thetag{1.3} above) one should impose some extra
limitations on a choice of Riemann surface $\Gamma$ and divisors
$D$ and $\hat D$ in addition to that of \cite{7}. Let us take
$\Gamma$ such that
\roster
\item branching points $\pm\lambda_1,\dots,\pm\lambda_g$ form
the set invariant under the action of the inversion
$\pi:\lambda\longrightarrow -\varkappa^6/\lambda$, where
$\varkappa$
is a positive constant.
\endroster
This case $\Gamma$ admits an involution $\pi:\lambda(\pi P)=
-\varkappa^6/\lambda(P)$ commuting with the involution $\sigma:\
\lambda(\sigma P)=-\lambda(P)$ and with the antiinvolution
$\tau:\ \lambda(\tau P)=\overline{\lambda(P)}$. Let us also choose
$D$ and $\hat D$ such that
\roster
\item[2] the divisor $\hat D$ is a map image $\hat D=\pi D$ of
the divisor $D$. One can check that this condition is compatible
with the previous restrictions for the choice of the divisors
\endroster
$$
\aligned
&D+\sigma D-P_\infty-P_0=C\\
&\hat D+\sigma \hat D-P_\infty-P_0=C
\endaligned\tag2.3
$$
where $\lambda(P_\infty)=\infty$, $\lambda(P_0)=0$ and $C$ is the
divisor of canonic class on $\Gamma$(see \cite{7}). \par
\proclaim{Lemma 1} Conditions \therosteritem{1} and
\therosteritem{2} above are enough for the finite-gap solutions
$v(x,t)$ and $\hat v(x,t)$ to be bound with the relationship
\thetag{2.2}.
\endproclaim
     One can obtain the proof of this lemma by comparing the
analytical properties (i.e. essential singularities  and poles)
of two Baker-Achiezer functions $\hat e(x,t,P)$ and
$e(-t/\varkappa^2 , -\varkappa^2 x,\pi P)$. \par
\head
3. The differential B\"acklund transformation.
\endhead
     Next step is to realize the relationship \thetag{1.4}.
According to \cite{8} each one-soliton solution $u(x,t)$ of the
equation \thetag{2.1} on a finite-gap background originate from
the appropriate one-soliton Baker-Achiezer function $\Psi(x,t,
P)$. This function has two exponential singularities at
$P_\infty$ and at $P_0$ of the same form as $e(x,t,P)$ and poles
forming the divisor $D^\circ+\Lambda+\Lambda^*$. Here $D^\circ$
is the divisor of degree $g$ defining one more finite-gap
solution of the equation \thetag{2.1} called the background
solution in \cite{8}. Two points $\Lambda$ and $\Lambda^*$
such that $\sigma\Lambda\neq\Lambda^*$ and $\lambda(\Lambda)=
-\Lambda(\Lambda^*)=\lambda_0$ form the discrete spectrum of the
soliton in question. Since $\Gamma$ is the three-fold covering
over $\lambda$-plane we can find one more point $\tilde\Lambda$
on $\Gamma$ in addition to $\Lambda$ and $\Lambda^*$ such that
$\lambda(\tilde\Lambda)=\lambda(\Lambda)=-\lambda(\Lambda^*)=
\lambda_0$. In terms of $\tilde\Lambda$ we can state the new
restrictions for the choice of $D$, $\hat D$ and $D^\circ$
taking the following divisors
$$
\aligned
&D^\circ+\Lambda+\Lambda^*+\tilde\Lambda-\hat D-3 P_\infty
\cong 0\\
&D^\circ+\Lambda+\Lambda^*+\sigma\tilde\Lambda-D-3 P_\infty
\cong 0
\endaligned\tag3.1
$$
to be linear equivalent to zero. This means that there are two
meromorphic functions $\hat\alpha(P)$ and $\alpha(P)$ whose
zeros and poles constitute the divisors \thetag{3.1}. One can
check that conditions \thetag{3.1} are compatible with
\thetag{2.3} and the same condition for $D^\circ$. From
\thetag{3.1} we derive the equivalence
$$
\hat D-D\cong\tilde\Lambda-\sigma\tilde\Lambda\tag3.2
$$
which is compatible with $\hat D=\pi D$ only if $\pi\tilde
\Lambda=\sigma\tilde\Lambda$. This means that $\tilde\Lambda$
is a stable point of the involution $\pi\circ\sigma$ and
therefore
$$
\lambda_0=\lambda(\tilde\Lambda)=\varkappa^3\tag3.3
$$
This case we write \thetag{3.2} as a restriction for the choice
of $D$
$$
\pi D-D\cong\tilde\Lambda-\sigma\tilde\Lambda\tag3.4
$$
Reality and non singularity conditions for the solutions
$v(x,t)$ and $\hat v(x,t)$ in terms of divisors $D$ and $\hat D$
(see \cite{8}) together with \thetag{3.4} fix the position of the
point $\tilde\Lambda$ on an invariant cycle $\alpha_0$ of the
antiinvolution $\tau$ passing trough $P_\infty$ and $P_0$. \par
\proclaim{Lemma 2} For the above defined meromorphic functions
$\alpha(P)$ and $\hat\alpha(P)$ normalized by the condition
$$
\alpha(P)\sim\hat\alpha(P)\sim\lambda(P)
\text{ as } P\to P_\infty
$$
their values at $P=P_0$
$$
\alpha(P_0)=-\lambda(\tilde\Lambda)=-\lambda_0\hskip 3em
\hat\alpha(P_0)=-\lambda(\sigma\tilde\Lambda)=\lambda_0
$$
are defined by the discrete spectrum $\lambda=\lambda_0=
\varkappa$ of the intermediate soliton solution $u(x,t)$ in
\thetag{1.4}.
\endproclaim
    \demo{Proof} Considering zeros and poles for meromorphic
function $\alpha(P)\alpha(\sigma P)$ and its normalizing
condition we get
$$
\alpha(P)\alpha(\sigma P)=-\frac{\omega^\circ(P)}{\omega(P)}
\left[\lambda^2(P)-\lambda^2_0\right]
\tag3.5
$$
where $\omega(P)$, $\hat\omega(P)$ and $\omega^\circ(P)$ are
abelian differential of the third kind defined by the divisors
\thetag{2.3} and $D^\circ+\sigma D^\circ-P_\infty-P_0$ of
canonic class on $\Gamma$ (see \cite{7}). Substituting $P=P_0$
in \thetag{3.5} gives $\alpha^2(P_0)=\lambda^2_0$. The sign of
$\alpha(P_0)$ then is defined by considering zeros and poles
on a cycle $\alpha_0$ for the function $\alpha(P)$ which is
real valued on that cycle. For $\hat\alpha(P_0)$ proof is quite
similar.\qed\enddemo
     Now using $\alpha(P)$ and $\hat\alpha(P)$ we slightly
modify the one soliton Baker-Achiezer function $\Psi(P)$ to
get the functions $\alpha(P)\Psi(P)$ and $\hat\alpha(P)
\Psi(P)$ whose poles form the divisors $3 P_\infty-D$ and
$3 P_\infty-\hat D$. Comparing the analytical properties of
$\hat\alpha(P)\Psi(P)$ and $e(P)$ we get the matrix relationship
$$
\alpha(P)\Psi(P)=\left[\Cal C\lambda(P)+\Cal D\right]
e(P)\tag3.6
$$
where $\Cal C$ is the upper-triangular matrix with unities in
the diagonal and $\Cal D$ is the lower-triangular matrix with
the following values in the diagonal
$$
\Cal D_{11}=\alpha(P_0) e^{v-u}\qquad
\Cal D_{22}=\alpha(P_0) e^{u-v}\qquad
\Cal D_{11}=\alpha(P_0) \tag3.7
$$
Similarly for the function $\hat\alpha(P)\Psi(P)$ we have
$$
\hat\alpha(P)\Psi(P)=\left[\hat\Cal C\lambda(P)+\hat\Cal D
\right]e(P)\tag3.8
$$
with the matrices of the similar shape and with diagonal elements
in the matrix $\hat\Cal D$ being of the form
$$
\hat\Cal D_{11}=\hat\alpha(P_0) e^{\hat v-u}\qquad
\hat\Cal D_{22}=\hat\alpha(P_0) e^{u-\hat v}\qquad
\hat\Cal D_{11}=\hat\alpha(P_0) \tag3.9
$$\par
\proclaim{Lemma 3} Equivalence \thetag{3.4} is enough for the
finite gap solutions of the equation \thetag{2.1} corresponding
to the divisors $D$ and $\pi D$ to be bound with the one soliton
solution of this equation via the B\"acklund transformations
\thetag{1.4}.
\endproclaim
     Matrix equations \thetag{3.6} and \thetag{3.8} are the
integral form of transformations \thetag{1.4}. Using the
equations
$$
\alignat2
&\partial_xe(P)=\bold L(v)e(P)&&\hskip 3em
 \partial_te(P)=\bold A(v)e(P)\\
&\partial_x\Psi(P)=\bold L(u)\Psi(P)&&\hskip 3em
 \partial_t\Psi(P)=\bold A(u)\Psi(P)
\endalignat
$$
of the direct spectral problem with Lax operators $\bold L$ and
$\bold A$ one can obtain the following equations for the matrix
$\Cal B=\Cal C\lambda+\Cal D$
$$
\partial_x\Cal B=\bold L(u)\Cal B-\Cal B\bold L(v)
\hskip 3em
\partial_t\Cal B=\bold A(u)\Cal B-\Cal B\bold A(v)
\tag3.10
$$
Taking into account the special form of the matrices $\Cal C$
and $\Cal D$ and \thetag{3.7} from \thetag{3.10}we derive the
further specialization for $\Cal B$
$$
\Cal B=\left[\matrix\format\l&\qquad\l&\qquad\l\\
\lambda+be^{v-u}&p\lambda-\lambda(u_x-v_x)&q\lambda e^{-u}\\
(q-b(u_t-v_t))e^{u+v}&\lambda+be^{u-v}&\lambda p\\
p&qe^{-v}&\lambda+b
\endmatrix
\right]
$$
where $b=\alpha(P_0)=-\lambda_0=-\varkappa^3$. Matrix $\hat
\Cal B=\hat\Cal C+\hat\Cal D$ has the similar structure derived
with use of \thetag{3.9}. Substituting the specialized form of
matrix $\Cal B$ into equation \thetag{3.10} one can extract
a lot of differential equations part of which forming the
differential form of B\"acklund transformation for the equation
\thetag{2.1} and others being their consequences. For the sake
of brevity we introduce the following notations:
$$
\aligned
&U=(u+v)/2\\ &h=e^{-U}\cosh(V)-e^{2U}
\endaligned\hskip 3em
\aligned
&U=(u-v)/2\\ &R=V_xV_t-h^2
\endaligned\tag3.11
$$

For $p(x,t)$ and $q(x,t)$ then we have
$$
\aligned
p&=2 e^{2U} R^{-1} (h V_x-b V^2_t)\\
q&=2 e^{2U} R^{-1} (b h V_t - V^2_x)
\endaligned
\tag3.12
$$
Moreover these functions are connected with $U(x,t)$ and
$V(x,t)$ via the differential equations
$$
\aligned
p_t&=2e^U\sinh(V)\\ q_x&=2be^U\sinh(V)
\endaligned\hskip 3em
\aligned
p_x&=U_xp-e^{-U}q\sinh(V)\\ q_t&=U_t-e^Up\sinh(V)
\endaligned\tag3.13
$$\par
\proclaim{Theorem 1} The differential form of the B\"acklund
transformation for the equation \thetag{2.1} consist of three
equations
$$
\align
b^{-1}V_x^3+&bV_t^3+e^{-2U}R^2-3hR-2h^3=0\tag3.14\\
&\aligned
 V_{xx}+2U_xV_x&=\frac32p_x\\
 V_{tt}+2U_tV_t&=\frac32b^{-1}q_t
 \endaligned\tag3.15
\endalign
$$
being after the substitution \thetag{3.13}, \thetag{3.12} and
\thetag{3.11} in \thetag{3.14} and \thetag{3.15} the set of
differential equalities in $u(x,t)$, $v(x,t)$ consistent with
the equation \thetag{2.1} and closed  in the sense  that it has
no functionally independent differential consequences with
the first and second order derivatives.
\endproclaim
     Though the B\"acklund transformation \thetag{3.14},
\thetag{3.15} is derived on a base of finite-gap solutions,
it is applicable to arbitrary solutions of the equation
\thetag{2.1} since the the proof of the theorem 1 require
nothing but direct calculations. \par
\head
4. On the extraction of the integrable boundary condition.
\endhead

     Let us consider the pair of B\"acklund transformations
\thetag{1.4} together with the nonlocal equality \thetag{2.2}.
Note that \thetag{2.2} is local when restricted on a moving
boundary $x=-t/\varkappa^2$. To deal with the stable boundary
$x=0$ we should introduce new variables by substituting
$\varkappa^{-1}(x-t)/2$ and $\varkappa(x+t)/2$ for $x$ and $t$
respectively. On doing that we find that the equation
\thetag{2.1} takes its initial form \thetag{1.2} and
\thetag{2.2} becomes \thetag{1.3}. As a consequence of
\thetag{1.3} we obtain the equalities
$$
\aligned
\hat v&=v\\ \hat v_{xt}&=-v_{xt}
\endaligned\hskip 3em
\aligned
\hat v_x&=-v_x\\ \hat v_{xx}&=v_{xx}
\endaligned\hskip 3em
\aligned
\hat v_t&=v_t\\ \hat v_{tt}&=v_{tt}
\endaligned\tag4.1
$$
which hold on a boundary $x=0$. Using the same notations
\thetag{3.11} except for
$$
R=v_x^2-V_t^2-h^2\hskip 3em \hat R=U_x^2-V_t^2-h^2
\tag4.2
$$
we can rewrite the equation \thetag{3.14} for the changed
variables $x$, $t$
$$
\aligned
&2V_x^3+6V_xV_t^2+e^{-2U} R^2-3hR-2h^3=0\\
&2U_x^3+6U_xV_t^2+e^{-2U}\hat R^2-3h\hat R-2h^3=0
\endaligned\tag4.3
$$
The second of these equations correspond to the second
transformation from \thetag{1.4}. Combining the equations
\thetag{3.15}
and \thetag{3.16} and changing variables $x$, $t$ in them for new
ones above we obtain the following two differential equations
$$
\align
&{\aligned
 &V_{xx}+V_{tt}+2U_xV_x+2U_tV_t=3he^U\sinh(V) R^{-1}V_x+\\
 &+3he^{2U}(U_xV_x+U_tV_t)R^{-1}+3e^U\sinh(V)(V_x^2+V_t^2)R^{-1}+\\
 &+3e^{2U}(U_xV_x^2+U_xV_t^2-2U_tV_xV_t)R^{-1}
 \endaligned}\tag4.4\\
&{\aligned
 &V_{xt}+2U_xV_t+2U_tV_x=-3he^U\sinh(V)R^{-1}V_t+\\
 &+3he^{2U}(U_xV_t+U_tV_x)R^{-1}+6e^U\sinh(V)R^{-1} V_xV_t+\\
 &+3e^{2U}(U_tV_x^2+U_tV_t^2-2U_xV_xV_t)R^{-1}
 \endaligned}\tag4.5
\endalign
$$
It is worth to note that these equations do not contain the
constant parameter $b=-\varkappa^3=-\lambda_0$ as well as the
equations \thetag{4.4} (see also \thetag{3.3} above). The second
B\"acklund transformation from \thetag{1.4} via \thetag{4.1} give
rise to the next two equations
$$
\align
&\aligned
 &V_{xx}+V_{tt}+2U_xV_x+2U_tV_t=-3he^U\sinh(V)\hat R^{-1}U_x+
 \\ &+3he^{2U}(U_xV_x+U_tV_t)\hat R^{-1}+3e^U\sinh(V)(U_x^2+
     V_t^2)\hat R^{-1}-\\
 &-3e^{2U}(V_xU_x^2+V_xV_t^2-2U_tU_xV_t)\hat R^{-1}
 \endaligned\tag4.6\\
&\aligned
 &2U_t+2V_xV_t+2U_tU_x=3he^U\sinh(V)\hat R^{-1} V_t+\\
 &+3he^{2U}(V_xV_t+U_tU_x)\hat R^{-1}+6e^U\sinh(V)\hat R^{-1}
 U_xV_t-\\
 &-3e^{2U}(U_tU_x^2+U_xV_t^2-2U_xV_xV_t)\hat R^{-1}
 \endaligned\tag4.7
\endalign
$$
which are also independent of $b=-\varkappa^3=-\lambda_0$. Now by
subtracting \thetag{4.4} from \thetag{4.6} and by subtracting
\thetag{4.5} from \thetag{4.7} we obtain two equations
containing the only derivative of the second order $v_{xt}=
U_{xt}-V_{xt}$. These two equations together with \thetag{4.4}
may be written in the form of four polynomial equations in the
following independent variables
$$
v_{xt}, v_x, v_t, e^v, u_x, u_t, e^u
$$
Excluding last three variables one can get the integrable
boundary condition  of the form \thetag{1.1} which is
automatically fulfilled for the finite-gap  solutions
$v(x,t)$ of the equation \thetag{1.2} constructed above
$$
R(e^v,v_x,v_t,v_{xt})\lowbar{x=0}=0
$$
Note that polynomial $R$ here is different from that of
\thetag{3.11} and \thetag{4.2}. It is not yet known to us.
The only obstacle that prevented us from getting the polynomial
$R$ in the explicit form is the enormous amount of calculations
needed to complete this task. We suppose this task may be of
interest as the touchstone for testing the computer algorithms
of polynomial algebra. \par

\enddocument
\end